**The Reconstructions of Konrad Zuse's Z3 Computer**

Raúl Rojas
Dept. of Mathematics & Statistics
University of Nevada Reno

What motivates engineers to reconstruct historical computers? Vintage Computer Festivals are held regularly in Europe and the United States, where early computers are presented to new generations of programmers and enthusiasts. This interest can be attributed, at least in part, to the central role computers have come to play in the modern industrial revolution. Historical computers have consequently acquired the status of cultural artifacts comparable to nineteenth-century steam engines, which are commonly preserved and exhibited in technical museums.

Such reconstructions are undoubtedly driven by engineers' curiosity and their desire to understand how these machines worked. At the same time, they help preserve our technical and cultural heritage.

A notable example is the Z3 computer developed by Konrad Zuse. Completed in 1941, it is now widely regarded as one of the pivotal artifacts in the history of computing [1]. The machine operated fully automatically, used binary floating-point numbers, and read its instructions from perforated film tape. The original, however, was destroyed during an air raid on Berlin in December 1943. This meant the loss not only of the machine itself but also of the most important material evidence of its operation. Only a small number of people had seen it working.

On June 16, 1941, Zuse filed a patent application with the German Patent Office for a "calculating device" under application number Z 26476 [2]. The application described the fundamental combination of a binary arithmetic unit, memory, and a control unit capable of automatically executing a program. After the war, Zuse continued with the application process. This led to a long and difficult patent evaluation in which he had to defend the novelty and inventive significance of his design against multiple objections.

Against this background, Konrad Zuse commissioned his own company, Zuse KG, to construct a functional replica of the Z3 in the early 1960s. The reconstruction was intended to make the design described in the patent specification tangible and technically comprehensible, thereby supporting both Zuse's patent claim and his claim to historical priority. Because the original machine and much of its documentation had been destroyed during the war, the replica also served as a form of technical evidence.

This first replica marked the beginning of a distinct history of Z3 reconstructions. Over the following decades, several further machines were built, differing considerably in their objectives, technical implementation, and degree of fidelity to the original. Some were intended primarily as museum exhibits, while others were constructed for historical research, technical analysis, or public demonstrations of relay computing. The following sections present these reconstructions in chronological order.

### The Deutsches Museum Reconstruction

The first replica of the Z3 was completed in 1962 by Zuse KG under the direction of Konrad Zuse himself (Fig. 1). It was first displayed publicly from August 26 to September 2, 1962, at the Interdata computer exhibition in Munich, held in parallel with the first World Congress of the International Federation for Information Processing (IFIP). In May 1968, the replica was transferred to the communications technology department of the Deutsches Museum in Munich, where it has remained on display for decades.

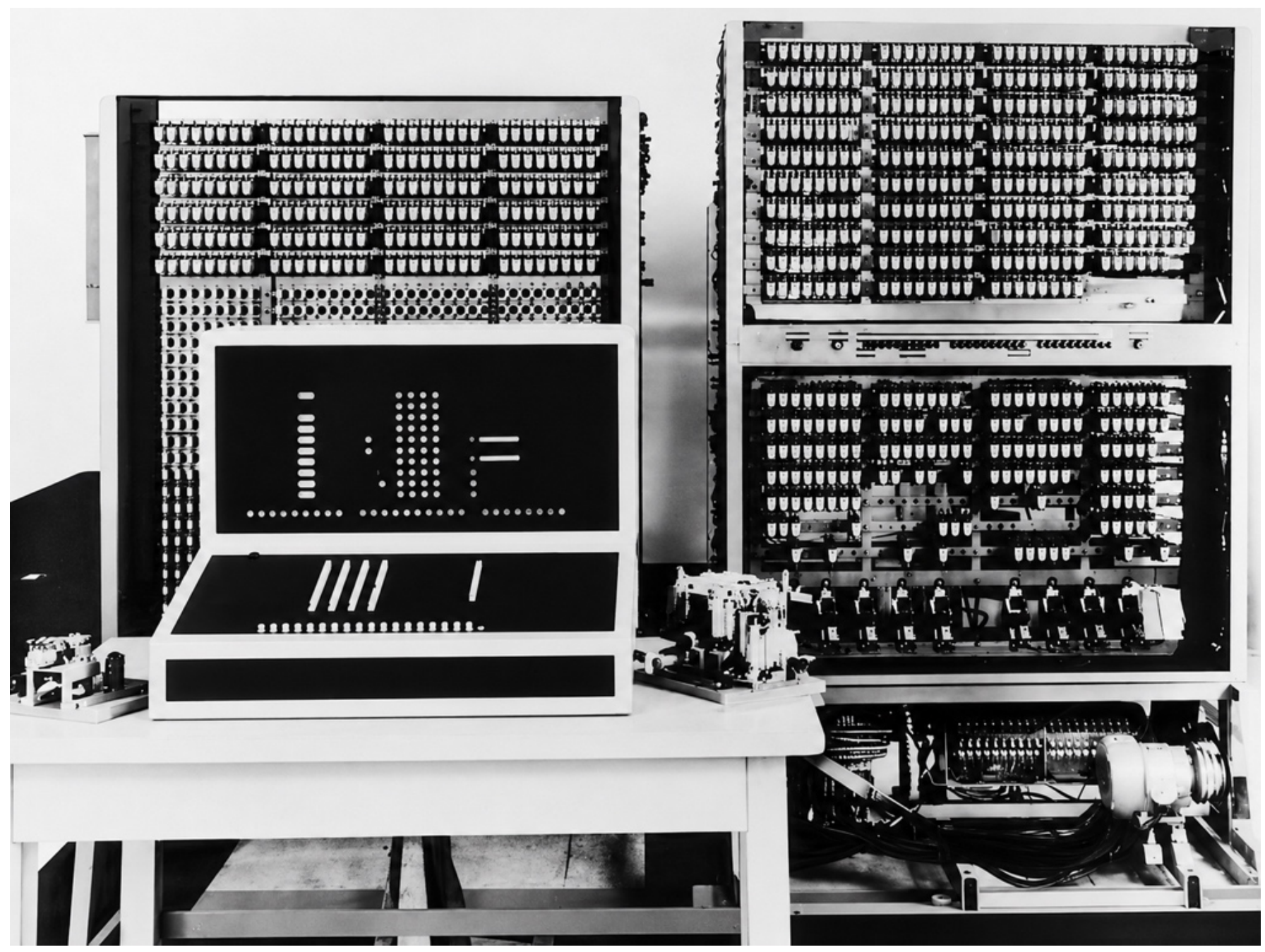

Fig. 1: The 1962 reconstruction of the Z3. The console is in front, the memory unit on the left, and the arithmetic unit on the right (Wikimedia Commons).

This machine, however, is not a faithful reconstruction in the modern sense. The original Z3 had been completely destroyed in December 1943; none of its components survived, and its external appearance was only incompletely documented. Zuse therefore had to rely on surviving circuit diagrams, a small number of photographs and sketches, and his own memory. The reconstruction used newer and smaller relays, and the spatial arrangement was modified slightly. It was smaller than the original and had only one memory cabinet, whereas the 1941 Z3 had two. The reconstruction therefore reproduced primarily the logical and functional operation of the Z3 rather than every structural and material detail of the original.

The replica itself, by contrast, is well documented. The archives of the Deutsches Museum contain the technical records produced by Zuse KG for its construction, including circuit diagrams, drawings, and other documents. The circuitry of this reconstruction can therefore be studied much more precisely than that of the lost 1941 machine.

## The Berlin Reconstruction of 2001

Another reconstruction of the Z3 was carried out under the direction of Raúl Rojas at the Free University of Berlin. Horst Zuse of the Technical University of Berlin arranged the funding for the project. On the basis of Konrad Zuse's patent application Z391, Rojas designed the block architecture of the memory and arithmetic unit [3]. Frank Darius developed the detailed relay circuits, which Georg Heyne then adapted to the available hardware. Wolfram Däumel, Lothar Schönbein, Torsten Vetter, and Cüneyt Göktekin also contributed to the project.

The machine was first demonstrated publicly on May 11, 2001, at the symposium "60 Years of Computer History—Konrad Zuse's Z3: 1941–2001," held at the Konrad Zuse Center for Information Technology in Berlin-Dahlem (Fig. 2). Its design followed the logical architecture of the Z3, but it was deliberately not intended as an externally faithful replica. Small LEDs made the flow of data visible and gave the machine a distinctly didactic character.

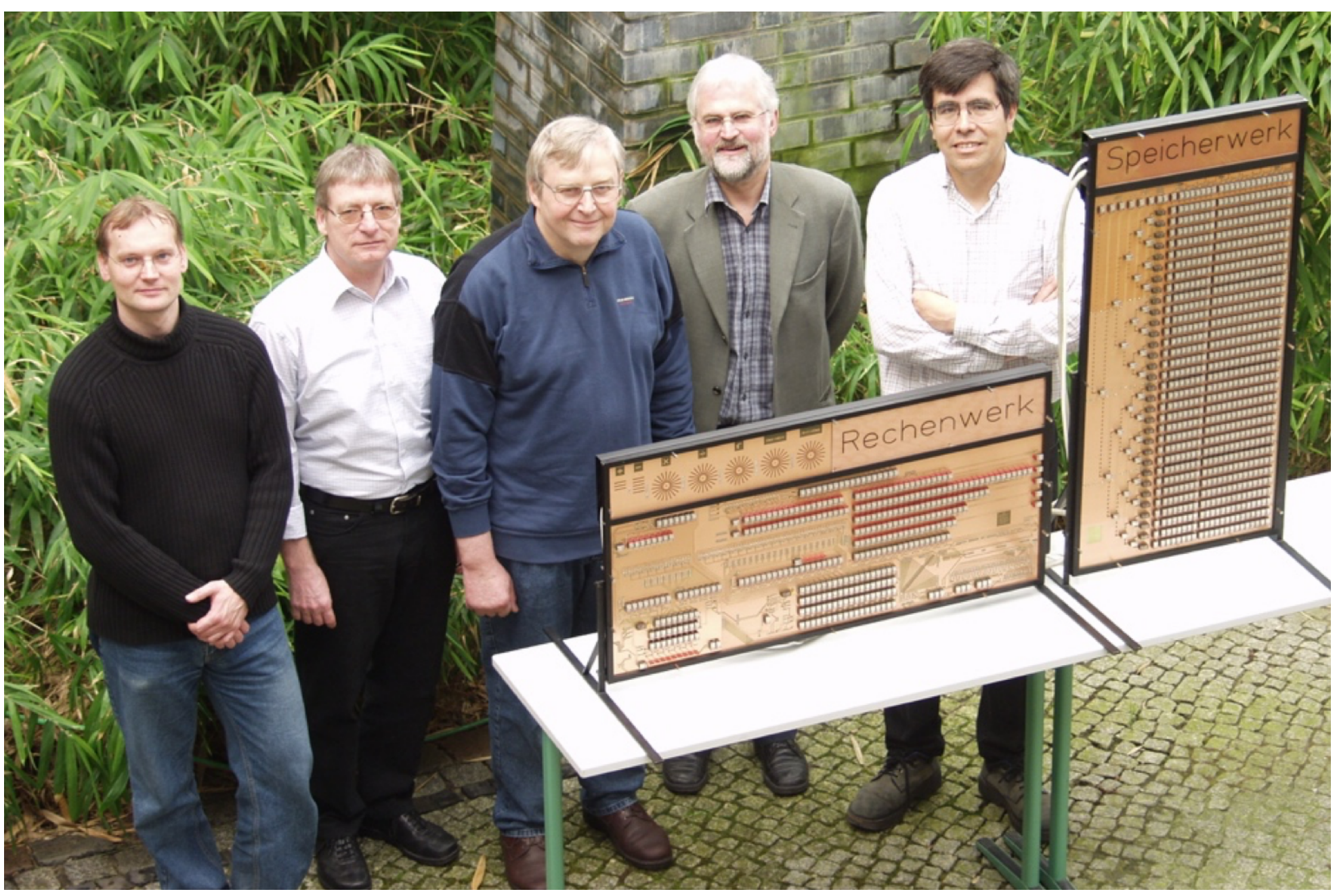

Fig. 2: The 2001 reconstruction of the Z3. Modern relays were interconnected by conductive tracks on circuit boards. The console and punched tape were simulated by means of a display. Photograph by R. Rojas.

Technically, this machine is a hybrid reconstruction. The memory and arithmetic unit are built with electromechanical relays, whereas modern digital components are used to control the execution of the arithmetic algorithms. The historical input and output console was not mechanically reproduced but replaced by a connected PC. The PC handles numerical input and output, machine operation, and the simulation of the punched program tape. The reconstruction therefore reproduces primarily the Z3's data path and arithmetic logic while using modern electronics and software for control and operation. The machine is now housed in the Konrad Zuse Museum in Hünfeld.

### Horst Zuse's Z3r

A further reconstruction was initiated by Horst Zuse in 2008 [4]. Known as the Z3r—short for "Z3 reconstructed"—the machine was intended to be presented in June 2010 to mark the centenary of Konrad Zuse's birth. It was built at approximately the original scale, with an arithmetic unit, two memory cabinets, and an operator panel modeled on the historical console (Fig. 3). Its first official public demonstration took place on June 26, 2010, at the Konrad Zuse Museum in Hünfeld. Work on the machine, including further modifications, continued until 2011.

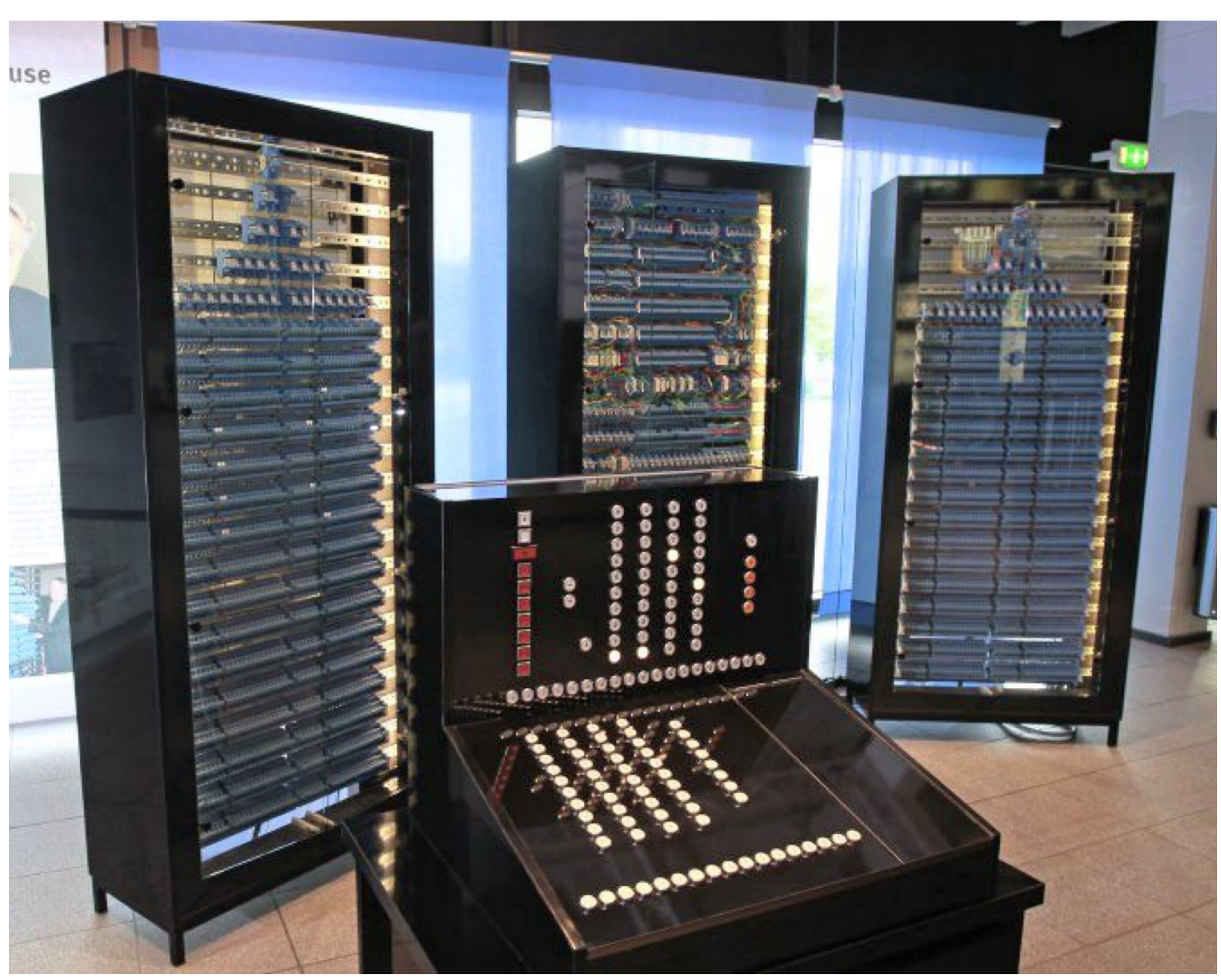

Fig. 3: Horst Zuse's Z3r at an exhibition at the Heinz Nixdorf MuseumsForum. Photograph by H. Zuse.

Horst Zuse used approximately 2,500 modern industrial relays manufactured by Finder. These are mounted on easily accessible rails, can be replaced individually, and are equipped with LEDs indicating their switching states. The historical stepping switches were replaced by modern time-delay relays.

The Z3r was designed especially for educational purposes. Clock control, exponent and mantissa arithmetic, shifting operations, normalization, and conversion between decimal and binary floating-point numbers can all be demonstrated separately. Beneath a flap in the control console are approximately sixty pushbuttons that allow individual components to be controlled down to the level of single bits. The machine was therefore intended not only to reproduce the external form and circuit logic of the Z3 but, above all, to make its operation directly visible to visitors, school pupils, and university students. It is now located in the German Museum of Technology in Berlin.

### Christof Traber's Reconstruction in Solothurn

The next reconstruction of the Z3 was built between 2021 and 2024 under the direction of the Swiss computer scientist Dr Christof Traber. Closely based on Konrad Zuse's 1941 patent application, it is now housed at the Enter Technikwelt Solothurn in Derendingen. The machine has only one memory unit, with a capacity of 31 words, while its control and arithmetic units are larger than those of the original Z3. Because of the generally wider spacing between the relays and the didactic layout of the arithmetic unit—with a clearly visible separation between the exponent and mantissa processing sections—the arithmetic and control units were not mounted one above the other, as in the original, but in two separate racks.

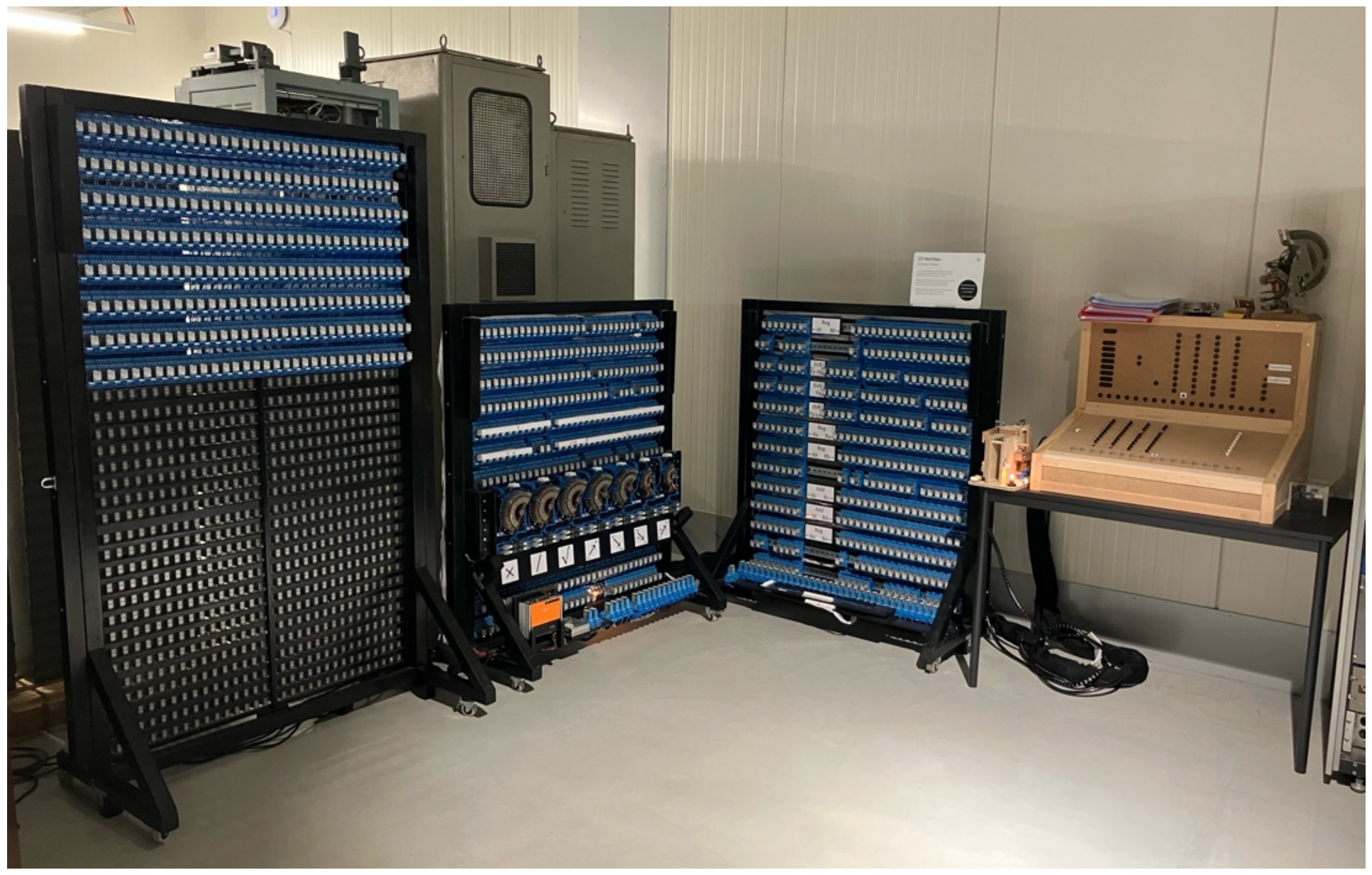

Fig. 4: Christof Traber's reconstruction in Switzerland. The three racks contain, from left to right, the memory, the control unit—with the clock generator in the two lowest rows of relays—and the arithmetic unit, with its clearly visible separation between exponent and mantissa processing. Beside them stands the console with the prototype punched-tape reader. Photograph by Ch. Traber.

The machine uses modern Finder relays. The dual-coil relays frequently used in the original, which are now difficult to obtain, were systematically replaced by a single standard relay and two diodes. Apart from these diodes, the reconstruction is fully electromechanical from a functional point of view. This also applies to the clock generator, in which a relay chain generates the various timing signals instead of a switching drum.

The prototype punched-tape reader currently operates with only four bits, thereby restricting the addressable memory range. It is planned to expand it to eight bits. The punched-tape reader is operated by a relay-based driver. To control the sequence of the arithmetic operations, Traber uses electromechanical stepping switches mounted at a 90-degree angle so that their rotary motion can be seen more clearly.

Traber's reconstruction is fully functional and makes it possible to follow the Z3's relay logic and binary floating-point arithmetic step by step. It is regularly demonstrated to the public at the Enter Technikwelt; demonstrations currently take place on the last Saturday of each month. The Solothurn machine is therefore one of the few operational Z3 reconstructions that are permanently accessible to the public and can regularly be seen in operation.

**Summary**

Several reconstructions of the Z3 have been built over the past decades, each pursuing different objectives. The replica constructed by Zuse KG was initially intended to support Konrad Zuse's patent claims and his claim to historical priority. Later reconstructions placed greater emphasis on technical research, museum presentation, or education. Some reproduce the logical operation of the Z3 using modern relays and electronic control systems, while others seek to approximate the original more closely through the use of historical components and electromechanical stepping switches.

None of these machines is therefore merely a copy of another. Each reconstruction has its own emphasis: visualizing computational processes, ensuring reliable public demonstrations, investigating the original circuitry, or approximating the material form of the historical machine. What they share is a commitment to preserving the operation and historical significance of the Z3 for future generations.

The Z3 is not merely a technical milestone but a major artifact of computer history. It is generally regarded as the world's first functional, freely programmable, fully automatic binary computer. Because the original machine of 1941 has not survived, the reconstructions are of particular importance. They make it possible to experience a machine that would otherwise be known only through patents, drawings, photographs, and written accounts.

Each reconstruction therefore contributes in its own way to preserving Konrad Zuse's achievement and the beginnings of the computer age. To paraphrase the conservationist William Morris, we are merely trustees for those who come after us. The reconstructions of the Z3 fulfill precisely this task: they preserve a lost piece of technical heritage and make it understandable and accessible to future generations.

**Acknowledgements**: The English grammar of this paper was improved using DeepL Write.